\documentclass[aps,prl,onecolumn,showpacs,superscriptaddress,groupedaddress]{revtex4}

\usepackage{amssymb}
\usepackage{color,graphicx}
\usepackage{amsmath}
\usepackage{amsbsy}
\usepackage{amsthm}
\usepackage{bbm}
\usepackage{bm}
\usepackage{epsfig}
\usepackage{lscape}
\usepackage{float}
\usepackage{subfigure}

\newcommand{\D}{\text{d}}
\newcommand{\com}[2]{\left[ {#1},{#2} \right]}

\begin{document}

\title{Upper bounds on spontaneous wave-function collapse models using millikelvin-cooled nanocantilevers }

\author{A. Vinante}
\email{anvinante@fbk.eu}
\affiliation{Istituto Nazionale di Fisica Nucleare (INFN), TIFPA, I-38123 Povo, Trento, Italy.}
\affiliation{Istituto di Fotonica e Nanotecnologie, CNR - Fondazione Bruno Kessler, I-38123 Povo, Trento, Italy.}

\author{M. Bahrami}
%\email{Mohammad.Bahrami@ts.infn.it}
\author{A. Bassi}
%\email{bassi@ts.infn.it}
\affiliation{Department of Physics, University of Trieste, Strada Costiera 11, 34014 Trieste, Italy}
\affiliation{Istituto Nazionale di Fisica Nucleare (INFN), Trieste Section, Via Valerio 2, 34127 Trieste,
Italy}

\author{O. Usenko}
\author{G. Wijts}
\author{T.H. Oosterkamp}
\affiliation{Leiden Institute of Physics, Leiden University, P.O. Box 9504, 2300 RA Leiden, The Netherlands}

\date{\today}

\begin{abstract}
Collapse models predict a tiny violation of  energy conservation, as a consequence of the spontaneous collapse of the wave function.  This property allows to set experimental  bounds on their parameters. We consider an ultrasoft magnetically tipped nanocantilever cooled to millikelvin temperature. The thermal noise of the cantilever fundamental mode has been accurately estimated in the range $0.03-1$ K, and any other excess noise is found to be negligible within the experimental uncertainty. From the measured data and the cantilever geometry, we estimate the upper bound on the Continuous Spontaneous Localization (CSL)  collapse rate in a wide range of the correlation length $r_C$. Our upper bound improves significantly previous constraints for $r_C>10^{-6}$ m, and partially excludes the enhanced collapse rate suggested by Adler. We discuss  future improvements.
\end{abstract}


\maketitle

Spontaneous wave function collapse models~\cite{GRW,CSL,collapse_review1,collapse_review2} have been proposed to conciliate the linear and deterministic evolution of quantum mechanics with the nonlinear and stochastic character of the measurement process. According to such phenomenological models, random collapses occur spontaneously in any material system, leading to a spatial localization of the wave function. The collapse rate scales with the size (number of constituents) of the system, in such a way as to produce rapid localization of any macroscopic system, while giving no measurable effect at the microscopic level, where conventional quantum mechanics is recovered. 

Here we consider the mass-proportional version of the Continuous Spontaneous Localization (CSL) model~\cite{CSL}, the most widely studied model, originally introduced as a refinement of the Ghirardi-Rimini-Weber (GRW) model \cite{GRW}.  At the density matrix level, the CSL model is described by a Lindblad type of master equation for the density matrix $\rho$, with the Lindblad term (projected on the $N$-particle subspace of the Fock space, in momentum representation) given by: 
\begin{align} \label{eq:1}
{\cal L}_{\text{\tiny CSL}} [\hat{\rho}(t)] =& \frac{(4\pi)^{\frac32}\,\lambda\,r_C^3}{2m_0^2}\,\sum_{i,j=1}^N\,m_im_j \int\frac{\D^3{\bf k}}{(2\pi)^3}\times \\\nonumber&~~~~~~ e^{-r_C^2{\bf k}^2}\, \com{e^{i\mathbf{k}\cdot\hat{\mathbf{x}}_i} }{ \com{e^{-i\mathbf{k}\cdot\hat{\mathbf{x}}_j}}{\hat{\rho}(t)}}, 
\end{align}
where $i$ and $j$ label the number of particles, $m_i$ and $\hat{\mathbf{x}}_i$ are the mass and position operator of particle $i$, and $m_0 = 1$ amu. This term causes the loss of quantum coherence, as an effect of the collapse process, and is responsible for the deviation from the standard quantum behavior.  

The CSL model is characterized by two phenomenological constants, a collapse rate $\lambda$ and a characteristic length $r_C$, which characterize respectively the intensity and the spatial resolution of the spontaneous collapse.
The standard conservative values suggested for CSL parameters are $\lambda \simeq  10^{-17}$ s$^{-1}$ and $r_C=10^{-7}$ m~\cite{GRW,CSL}. 
A strongly enhanced value for the collapse rate has been suggested by Adler \cite{adler}, motivated by the requirement of making the wave-function collapse effective at the level of latent image formation in photographic process. The values suggested by Adler are $\sim  10^{9 \pm 2}$ times larger than standard values at $r_C=10^{-7}$ m, and $\sim  10^{11 \pm 2}$ times larger at $r_C=10^{-6}$ m.

The direct effect of collapse models is to destroy quantum superpositions, resulting in a loss of coherence in interferometric matter-wave experiments~\cite{exp_MW,exp_MW2,exp_MW3}. Recently, non-interferometric tests have been proposed, which promise to set stronger bounds on these models~\cite{collett, adler2005, bassi2005,bassi,nimmrichter,diosi,goldwater}. Among such tests, the measurement of heating effects in mechanical systems, a byproduct of the collapse process, seems particularly promising~\cite{bassi,nimmrichter,diosi,goldwater}. In this work, we establish for the first time an experimental upper bound on the CSL collapse rate $\lambda$, by  accurate measurements of the mean energy of a nanocantilever in thermal equilibrium at millikelvin temperatures. This bound is found to be 2 orders of magnitude stronger than that set by matter-wave interferometry~\cite{ma, kl, mwi} for $r_C = 10^{-7}$m, and in general is the strongest one for $r_C>10^{-6}$ m.

%, and can be directly compared with other bounds set by spontaneous emission of X-ray \cite{curceanu} and matter-wave interferometry~\cite{ma, kl, mwi}. In the latter case, in particular, our bound is about 2 order of magnitude stronger for $r_C = 10^{-7}$m.

%%%%%%%%%%%%%%%%%%%%%%%%%%%%%%%%%%%%%%%%%%%%%%%%%%%%%%%%%%%%%%%%%%%%%%%%%%%%%%%%%%%%
\noindent{\it Theoretical model -- }
%%%%%%%%%%%%%%%%%%%%%%%%%%%%%%%%%%%%%%%%%%%%%%%%%%%%%%%%%%%%%%%%%%%%%%%%%%%%%%%%%%%
The detection of CSL-induced heating in realistic optomechanical systems has been extensively discussed in the recent literature~\cite{bassi, nimmrichter, diosi, goldwater}. Here we summarize the main steps. We consider a mechanical resonator in equilibrium with a phononic thermal bath at temperature $T$. When the spatial motion of the resonator is smaller than $r_C$, as in our experiment ($|\Delta x|\sim10^{-9}\,$m), Eq.~\eqref{eq:1} can be Taylor expanded. In the case of a rigid body, the Lindblad effect on the center-of-mass motion becomes~\cite{bassi,nimmrichter}:
\begin{equation}\label{1D-master}
{\cal L}_{\text{\tiny CSL}} [\hat{\rho}_{\text{\tiny CM}}(t)] = -\eta\, [\hat{q}, [\hat{q}, \hat{\rho}_{\text{\tiny CM}}(t)]],
\end{equation}
$\hat{q}$ being the position operator of the center-of-mass, and
\begin{align}\label{eta}
\eta &=\frac{(4\pi)^{\frac32}\,\lambda\,r_C^3}{m_0^2}\,
\int\frac{\D^3{\bf k}}{(2\pi)^3}\,k^2_z\,e^{-{\bf k}^2r_C^2}\,|\tilde{\varrho}({\bf k})|^2
\end{align}
with ${\bf k}=(k_x,k_y,k_z)$, $\tilde{\varrho}({\bf k})=\int\D^3{\bf x}\,e^{i{\bf k}\cdot{\bf r}}\,\varrho({\bf r})$ and $\varrho({\bf r})$ the mass density of the oscillator. 
The motion is only in one spatial direction which we set as $z$-axis. The effect of the Lindblad term in Eq.\eqref{1D-master} can be mimicked by adding the stochastic potential $V(t)=-\hbar \,w_t\,\sqrt{\eta}\,\hat{q}$ to the Hamiltonian of the system, and then taking the stochastic average $\mathbb{E}(\cdot)$. Here, $w_t$ is a white noise, with zero average and delta-correlation function: $\mathbb{E}(w_{t}w_{s})=\delta(t-s)$. Accordingly, the Hamiltonian takes the following form:
\begin{equation} \label{eq:h}
\hat{H}=\frac{1}{2m}\hat{p}^2+\frac12m\omega_0^2\,\hat{q}^2 - \hbar \,w_t\,\sqrt{\eta}\,\hat{q},
\end{equation}
and the corresponding Heisenberg equations of motion, where we add a term describing the coupling of the oscillator with a phononic bath, are:
\begin{align}\label{QLE}
\begin{aligned}
\partial_t\hat{q}&=\hat{p}/m\,,\\
\partial_t\hat{p}& = -m\omega_0^2\hat{q} + \hbar\sqrt{\eta}\,w_t - \gamma_m\,\hat{p} + \hat{\xi}(t)\,,
\end{aligned}
\end{align}
where the stochastic operator $\hat\xi(t)$ describes the Brownian-motion induced by the phononic bath, and $\gamma_m$ is its friction constant. The autocorrelation of $\hat\xi(t)$, after tracing over all phononic modes, is given by~\cite{mauro,mauro2} $\mathbb{E}(\langle\hat \xi(t)\hat \xi(s)\rangle)=\frac{\hbar m\gamma_m}{2\pi}\int^{\infty}_{-\infty}\D\omega\,\omega e^{-i\omega(t-s)}(\coth(\beta\omega)+1)$, and $\beta=\hbar/(2k_BT)$ with $k_B$ the Boltzmann constant and $T$ the temperature of the phononic bath. Notice that in the high-temperature limit, one recovers the white noise relation, i.e. $\mathbb{E}(\langle\hat \xi(t)\hat \xi(s)\rangle)=2m\gamma_mk_BT\,\delta(t-s)$.

The physical quantity that is estimated in the experiment is the spectrum ${\cal S}_q(\omega)$, i.e. the Fourier transform of the two-time correlation function of the oscillator's position: ${\cal S}_q(\omega)=\int_{-\infty}^{+\infty}\D\tau\,e^{-i\omega\tau}\,
\mathbb{E}(\langle\hat{q}(t)\hat{q}(t+\tau)\rangle)$. The area covered by ${\cal S}_q(\omega)$ is a measure of the variance of $\hat{q}$, which is proportional to the mean energy, or equivalently the temperature, of the mechanical resonator.

Standard calculation~\cite{mauro,mauro2} leads to the following result, which holds in the high temperature limit:
\begin{equation}\label{eq:S_w_3}
{\cal S}_q(\omega)=\frac{\hbar}{4m\omega_0}\,\frac{2\gamma_mk_BT/\hbar\omega_0+\eta(\hbar/m\omega_0)}
{\left(\omega-\omega_0\right)^2+\gamma_m^2/4}.
\end{equation}
which clearly shows that CSL increases the spectrum, implying that the the mean energy is higher than what standard quantum mechanics predicts.

Given Eq.~\eqref{eq:h}, the equilibrium energy $\mathbb{E}(\langle\hat{H}\rangle)$ can be easily expressed in terms of the spectral density ${\cal S}_q(\omega)$ and ${\cal S}_p(\omega)$ of the oscillator's position and momentum~\cite{mauro,mauro2}.  Eq.\eqref{QLE} gives: $\hat{p}(\omega)=-im\omega\hat{q}(\omega)$, implying that ${\cal S}_p(\omega)=m^2\omega^2{\cal S}_q(\omega)$, and by using Eq.~\eqref{eq:S_w_3}, we arrive at the expression: $
\mathbb{E}(\langle\hat{H}\rangle)=k_B T+ \hbar^2 Q \eta / 2 m\omega_0 $,    
where $Q = \omega_0/\gamma_m$ is the quality factor. One arrives at the same result also by directly solving the CSL master equation~\cite{diosi,goldwater}.
Thus, the experimental signature of CSL is a slight temperature-independent violation of the equipartition theorem.
We can express the excess energy as a temperature increase: 
\begin{equation} \label{temp}
\Delta T_{\mathrm{CSL}} =  \frac{\hbar^2 Q}{2 k_B m \omega_0} \eta.
\end{equation}
%This difference in temperature is what is measured in the experiment. 

We still have to estimate $\eta$, which depends on the geometry of the system and on the two phenomenological CSL parameters, as given by Eq.~\eqref{eta}. Our experiment is based on a ultrasoft silicon cantilever, with length $L=100$ $\mu$m, width $w=5$ $\mu$m and thickness $d=0.10$ $\mu$m (Fig. 1(a)). A ferromagnetic microsphere based on a neodymium-iron-boron alloy (density $\varrho_s=7430$ kg/m$^3$) with diameter $2 R=4.5$ $\mu$m is attached to the free end of the cantilever (density $\varrho_c=2330$ kg/m$^3$) and is used for displacement detection, as described below. 

Finding $\eta$ is not straightforward, both because of the non trivial geometry of the system, and because the motion of the cantilever is not rigid. In the Supplementary Note we show in detail how to compute $\eta$, which then defines $\Delta T_{\mathrm{CSL}}$.
% Based on the geometry and materials of the cantilever, we can calculate $\eta$ and $\Delta T_{\mathrm{CSL}}$ through Eqs. (\ref{eta}) and (\ref{temp}).
% As derived explicitly in the Supplementary Note, the total effect includes a term associated with the sphere and one associated with the cantilever. In addition, a correlation term appears, due to the fact that the two bodies are close to each other. For the sphere an exact formula is available, while for the cantilever we derive an approximate expression based on considering the modal shape of the cantilever. 
Fig. 1b shows the calculated CSL-induced overheating due to cantilever and microsphere and the total one, as a function of $r_C$, assuming the standard collapse rate $\lambda=2.2 \times 10^{-17}$ s$^{-1}$ \cite{CSL}. The cantilever contribution is significant only for $r_C<10^{-7}$ m, while for $r_C>10^{-7}$ m the microsphere contribution becomes largely dominant. The larger effect of the microsphere is explained by the dependence of $\eta$ on the square of the density. The total overheating peaks at $r_C=1.4$ $\mu$m, of the order of the microsphere radius.

\begin{figure}[!ht]
\includegraphics{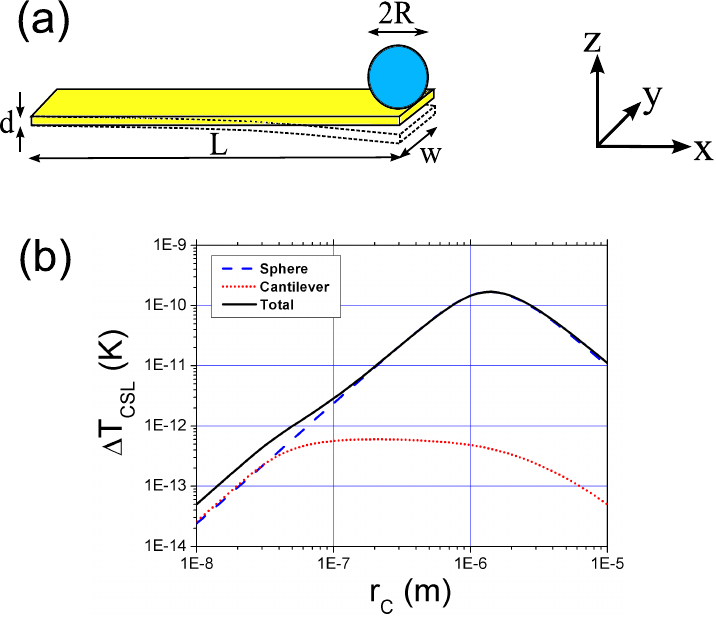}
\caption{(Color online). (a) Scheme of the mechanical resonator. We focus on the fundamental bending mode of a high aspect ratio nanocantilever with length $L$, width $w$ and thickness $d$. A ferromagnetic microsphere with radius $R$ is attached to the free end of the cantilever. (b) Calculated CSL-induced heating $\Delta T_{\mathrm{CSL}}$ of the cantilever fundamental mode as function of $r_C$. The total effect (solid line) includes two terms associated respectively to the microsphere (dashed line)and to the cantilever (dotted line), as well as a correlation term. Because of higher density, the contribution of the sphere is largely dominant for $r_C>10^{-7}$ m.  }  \label{fig1}
\end{figure}

%%%%%%%%%%%%%%%%%%%%%%%%%%%%%%%%%%%%%%%%%%%%%%%%%%%%%%%%%%%%%%%%%%%%%%%%%%%%%%%%%%%%
\noindent{\it Experimental Results -- }
%%%%%%%%%%%%%%%%%%%%%%%%%%%%%%%%%%%%%%%%%%%%%%%%%%%%%%%%%%%%%%%%%%%%%%%%%%%%%%%%%%%
Details on the detection scheme were already reported in Ref. \cite{usenko}. A SQUID current sensor is used to detect the motion of the magnetic particle on the cantilever via a superconducting pick-up loop. The cantilever chip is clamped above the superconducting detection coil by means of a brass spring, which also provides thermal contact to the thermal bath. The cantilever-coil setup is enclosed in a superconducting shield and thermally anchored to the mixing chamber of a dilution refrigerator. The temperature is monitored by a Speer resistive thermometer, calibrated against a high accuracy superconducting fixed-point reference device \cite{HDL}.  

\begin{figure}[!ht]
\includegraphics{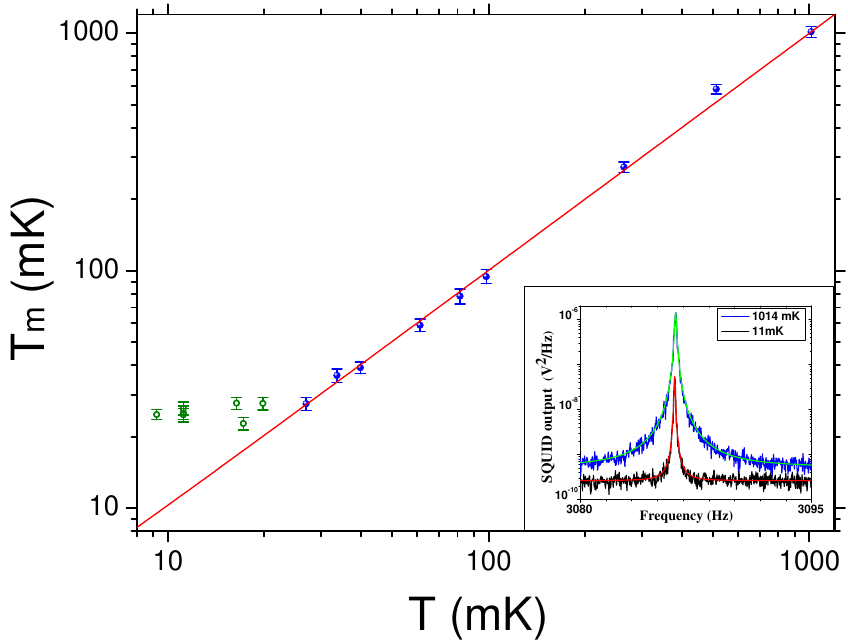}
\caption{(Color online). Main panel: cantilever mode temperature $T_m= E/ k_B$ as function of the bath temperature $T$. Data in the saturation region $T<25$ mK are excluded by the data analysis. Straight line represents the best fit with $T_m=\alpha T + T_0$. Inset: examples of acquired averaged spectra at two representative temperatures $T=11$ mK and $T=1.01$ K, with the respective best fit with a Lorentzian curve. Figures reproduced from Ref.~\cite{usenko}. }  \label{TvsT}
\end{figure}

The resonant frequency of the cantilever is $f_0=3084$ Hz and the quality factor measured with the ringdown method is $Q=38 \times 10^3$. Measurements of the mean energy of the cantilever mode, or equivalently the effective mode temperature $T_m$, were performed as a function of bath temperature $T$ in the range from $10$ mK up to $1$ K. The power spectrum of the SQUID-detected signal is acquired with a resolution of $0.02$ Hz, and at least 20 spectra are averaged for each point. The spectrum is well fitted by a Lorentzian peak associated to the cantilever motion incoherently superimposed on the SQUID white noise, as seen from the examples shown in the inset of Fig. \ref{TvsT}. An absolute calibration procedure has been developed to convert the area under the Lorentzian peak inferred from the fit into the mean energy $E$ of the cantilever mode. Details on the calibration procedure can be found in Refs. \cite{usenkosuppl,sashathesis}.

Fig.~\ref{TvsT} shows the measured cantilever mode temperature as function of the bath temperature. We have divided the dataset in two regions. For $T>25$ mK and up to the maximum temperature $\sim 1$ K, the data follow remarkably well the expected equipartition behaviour. In particular, the parameter-free equipartition curve $T_m=T$ fits the experimental data well, indicating that the cantilever is well thermalized and is actually behaving as a primary thermometer. Furthermore, we can infer that the systematic errors in the calibration and in the temperature measurement are negligible within the error bar. A linear fit with variable slope $\alpha$ gives $\alpha=\left( 1.03\pm 0.03 \right)$, indicating that the calibration systematic error is of the order 3$\%$ or less. 

At bath temperatures lower than $25$ mK, $T_m$ is found to saturate at an effective value $T_m=\left( 25 \pm 1 \right) $ mK. As discussed in Ref. \cite{usenko}, the saturation is consistent with an unknown effective heat leak to the cantilever on the order of $100$ aW. The sharpness of the saturation is typical at millikelvin temperature and is caused by the strong temperature dependence of the limiting thermalization mechanisms. For instance the heat conductivity of silicon or other thermal boundary resistances are expected to scale as $T^n$ with $n \sim 3 $. As a consequence, the cantilever mode temperature rapidly approaches the expected linear behaviour as soon as the temperature is increased above the saturation value.

The low temperature saturation cannot be attributed to CSL-induced heating, which would rather appear as a positive non-zero intercept of the measured data in the linear part. To set an upper bound on a possible CSL heating, we have to determine the maximum positive intercept consistent with the subset of experimental data following a linear behaviour. To this end, we perform a linear fit of the data above $25$ mK, with slope $\alpha$ fixed to 1 and the intercept $T_0$ as free parameter. The fit yields $T_0=\left( 0.28 \pm 1.18 \right)$ mK, with $\chi^2=1.20$. We may directly use this estimate to infer an upper limit at a given confidence level. However, one needs to be cautious when inferring upper limits on a quantity that is physically allowed to be only positive. Here, we adopt the Feldman-Cousins approach \cite{feldman}, which has been proposed precisely to address this kind of problems and to overcome possible misinterpretations of the confidence interval. In particular, we assume that the measured value $T_0$ provides an experimental estimation of the true value $\Delta T_{\mathrm{CSL}}>0$ of a positive CSL heating effect. Therefore, $T_0$ and $\Delta T_{\mathrm{CSL}}$ play the roles of $x$ and $\mu$ of Feldman-Cousins \cite{feldman}. The standard procedure for a Gaussian-distributed estimation $T_0$ provides then the upper limit $\Delta T_{\mathrm{CSL}} \leq 2.5$ mK at $95\%$ confidence level.

We have also performed the same procedure starting with a linear fit with both slope and intercept as free parameters. In this case, besides the slope $\alpha=1.03 \pm 0.03$, we obtain a slightly different estimate of the intercept $T_0=\left( -1.09 \pm 1.77 \right)$ mK. Nonetheless the final upper limit $\Delta T_{\mathrm{CSL}} \leq 2.4$ mK at $95 \%$ confidence level is essentially unchanged.

\begin{figure}[!t]
\includegraphics{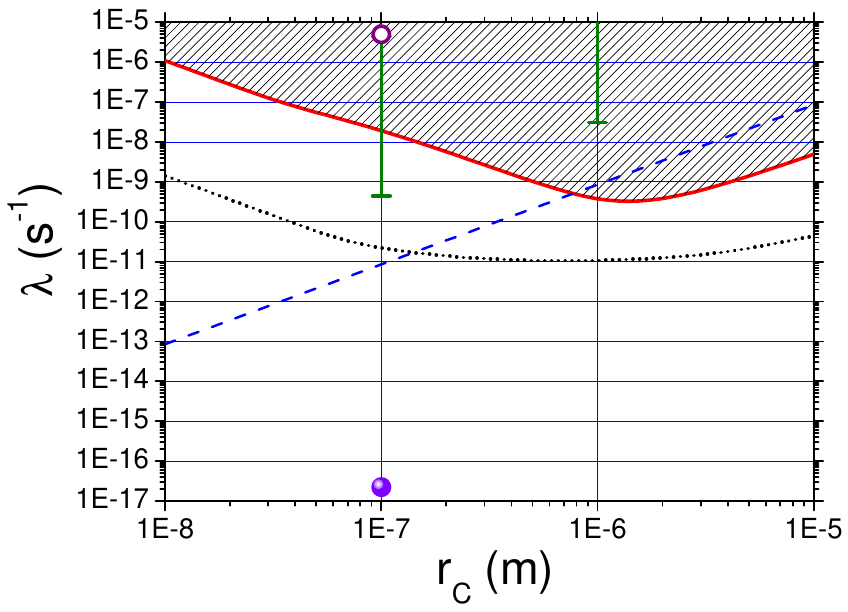}
\caption{(Color online). Exclusion plot in the $\lambda - r_C$ plane based on our experimental data, compared with the best experimental upper bounds reported so far and with the proposed theoretical lower bounds. Continuous (red) curve: upper limit on the CSL collapse rate $\lambda$, as function of the characteristic length $r_C$. The region above the curve is excluded at $95 \%$ confidence level. Dashed (blue) curve: upper limit on $\lambda$ from spontaneous X-ray emission \cite{curceanu}. Dotted (black) line: foreseen upper limit from the proposed future upgraded setup (see text). Hollow (purple) circle: best upper limit on $\lambda$ from matter-wave interferometry at $r_C=10^{-7}$ m~\cite{ma, kl, mwi}. Filled (violet) circle: conservative lower bound on CSL parameters according to Ghirardi {\it et al.} \cite{CSL}. Green bars: optimistic lower bounds on $\lambda$ at $r_C=10^{-7}$ m and $r_C=10^{-6}$ m as suggested by Adler \cite{adler}.  }   \label{upperlimit}
\end{figure}

%%%%%%%%%%%%%%%%%%%%%%%%%%%%%%%%%%%%%%%%%%%%%%%%%%%%%%%%%%%%%%%%%%%%%%%%%%%%%%%%%%%%
\noindent{\it Discussion -- }
%%%%%%%%%%%%%%%%%%%%%%%%%%%%%%%%%%%%%%%%%%%%%%%%%%%%%%%%%%%%%%%%%%%%%%%%%%%%%%%%%%%
Let us connect our experimental result to the CSL model.
By using Eq.~\eqref{temp} giving the expected CSL heating, which is a function of the collapse rate $\lambda$ and the correlation length $r_C$, and the measured upper limit on $\Delta T_{\mathrm{CSL}}$ discussed before, we can draw the exclusion plot shown in Fig. \ref{upperlimit}. The dashed region is excluded at $95\%$ confidence level.

In Fig. \ref{upperlimit} our upper limit is compared with the best one reported so far in literature, obtained by X-ray spontaneous emission experiments \cite{curceanu}. To allow for a full comparison we have extended the upper limit, reported only for $r_C=10^{-7}$ m in~\cite{curceanu}, to the full $r_C$ range. This is done by taking into account that CSL-induced X-ray emission scales as $r_C^{-2}$. We have also reported the upper bound coming from matter-wave interferometry~\cite{ma, kl, mwi}, $\lambda \leq 5.0 \times 10^{-6}$ s$^{-1}$ for $r_C = 10^{-7}$ m. Fig. \ref{upperlimit} shows also the conservative CSL lower bound according to the original paper of Ghirardi {\it et al.}~\cite{CSL} and the lower bounds suggested by Adler, based on the analysis of the latent image formation in photography~\cite{adler}. 

At the conventional length $r_C=10^{-7}$ m, our upper limit is still 3 orders of magnitude away from the limit set by X-ray emission, but provides an improvement over the X-ray limit at $r_C>10^{-6}$ m. It also improves the bound coming from matter wave interferometry by 2 orders of magnitude.

Compared with theoretical predictions, our limit is still 9 and 7 orders of magnitude far from the conservative collapse rate proposed by Ghirardi et al \cite{CSL} at $r_C=10^{-7}$ m and $r_C=10^{-6}$ m respectively. However, it compares favourably with Adler predictions. We remark that Adler intervals are lower bounds on CSL collapse rate. Our upper limit is thus ruling out Adler predictions completely at $r_C \geq  3\times 10^{-7}$ m, and partially at the conventional CSL length $r_C =  10^{-7}$ m. 

Despite Adler's lower bounds are already strongly excluded by X-ray experiments, our result is still significant because of the very different timescale involved. In fact, X-ray experiments probe the collapse field at very high frequency $\sim 10^{18}$ Hz, while we probe the collapse field at low frequency $\sim 1$ kHz, so that the two approaches are complementary. Moreover, it has been suggested that the limits inferred by X-ray emission could be evaded by assuming a high frequency cutoff on the collapse field spectrum \cite{adler}. In contrast, the timescale of our experiment is comparable to that assumed by Adler when analysing the process of latent image formation, which led to his enhanced lower bound. Therefore our data imply that Adler's proposal is ruled out, at least for $r_C > 3 \times 10^{-7}$ m, even under the assumption of non-white noise spectrum.

We conclude with an outlook towards future improvements of our results. First, we consider an upgraded setup with available technology. Single crystal diamond cantilevers with thickness $0.6$ $\mu$m have been recently demonstrated, with very high quality factors approaching 10$^7$ at 100 mK \cite{degen}. Combining such a device with a high density mass-load (we choose as an example a FePt film with size $40\times 12 \times 0.2$ $\mu$m \cite{FePt}) and assuming to be still able to detect a temperature excess of $\sim $1 mK, we obtain the dotted curve in Fig. \ref{upperlimit}. This would improve by 2-3 orders of magnitude the upper limit obtained in this work. Larger improvements towards the Ghirardi limit can be conceived, based either on devices operating at much lower frequency, such as torsion microbalances and micropendula \cite{matsumoto}, or on optically or magnetically levitated microparticles \cite{goldwater,romero}.

%%%%%%%%%%%%%%%%%%%%%%%%%%%%%%%%%%%%%%%%%%%%%%%%%%%%%%%%%%%%%%%%%%%%%%%%%%%%%%%%%%%
The experiment was financially supported by ERC and by the EU-project Microkelvin. AV acknowledges support from INFN. MB and AB acknowledge financial support from EU project NANOQUESTFIT, The John Templeton Foundation (grant n. 39530), University of Trieste (FRA 2013) and INFN. We thank W. Bosch for the high-accuracy temperature calibration, C. Curceanu, K. Piscicchia, S. Donadi, H. Ulbricht, M. Paternostro and M. Toros for useful discussions.

%%%%%%%%%%%%%%%%%%%%%%%%%%%%%%%%%%%%%%%%%%%%%%%%%%%%%%%%%%%%%%%%%%%%%%%%%%%%%%%%%%%

\section{Supplementary Note}

\setcounter{equation}{0}
\setcounter{figure}{0}
\renewcommand{\theequation}{S.\arabic{equation}}
\renewcommand{\thefigure}{S\arabic{figure}}

\section{Cantilever modal shape, effective mass and rigid cuboid approximation}
The motion along the $z$ axis of a given point of the cantilever, within one of its flexural modes (see Fig. 1(a) of main text), can be described by the displacement function:
\begin{equation}
z\left( {x,t} \right) = A\left( x \right) q \left( t \right).
\end{equation}
Here, $q \left( t \right)$ is the modal coordinate and $A\left( x \right)$ defines the x-dependent modal shape. We set $x=0$ and $x=L$ respectively as the clamped end and the free end of the cantilever. By definition $A\left( 0 \right)=0$, while $A\left( L \right)$ depends on a normalization factor. 
We choose the normalization $A \left( L \right) =1$, so that $q \left( t \right)$ coincides with the displacement of the free end of the cantilever. Notice that this coincides also with the displacement of the microsphere, which is the experimentally measured quantity.

\begin{figure}[ht!]
\includegraphics[width=12cm]{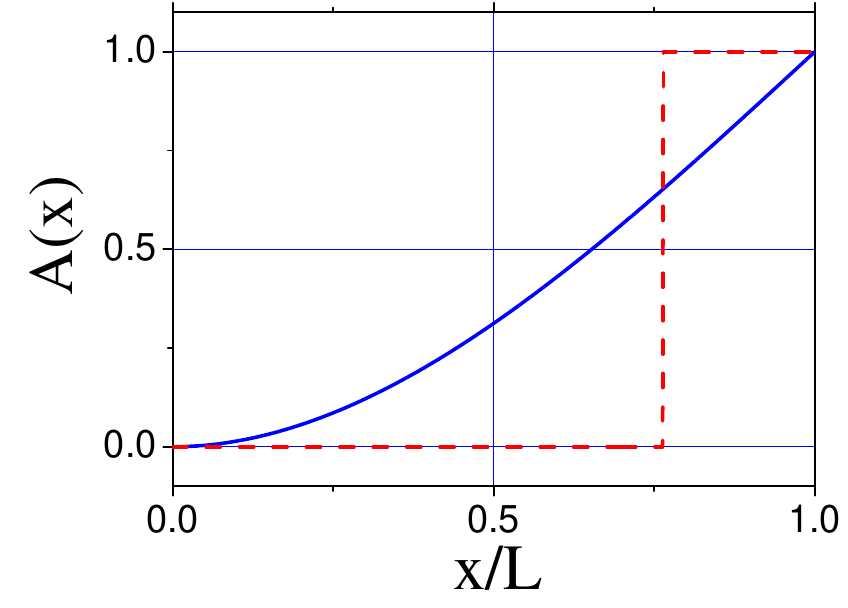}
\caption{Modal shape of the cantilever fundamental mode. Continouous blue line: Modal shape calculated using the exact analytical expression based on the Eulero-Bernoulli approximation. Dashed red line: unit step function approximation of the modal shape, with same effective mass of the exact modal shape. This approximation is used to estimate the cantilever contribution to the CSL force noise.}  \label{fig:modeshape}
\end{figure}

Analytical expressions of the modal shape can be calculated in the Eulero-Bernoulli approximation and can be found in many textbooks on elastic bodies. We follow the procedure described in Ref. \cite{cantilever}, which provides the modal shape of the flexural modes of a mass-loaded cantilever. In particular, the modal shape of the fundamental mode of the cantilever used in this work is shown in Fig.~\ref{fig:modeshape}.

A quantity which is relevant to this work is the effective mass $m_e$ of the cantilever mode, as seen from its free end.~This is defined by the expression $m_e=\beta m_c$ where $m_c=\varrho_c L w d$ is the physical mass and:
\begin{align}
\beta  = \frac{1}{L}\int\limits_0^L {A^2\left( x \right) } dx \approx 0.236.
\end{align}
It is straightforward to verify that the total kinetic energy within the mode vibration is given by $T_m=m_e v^2/2$, where $v=\dot q$ is the velocity of the free end. In other words, $m_e=\beta m_c$ is the effective mass of the cantilever if we choose to describe it as a mass rigidly oscillating with amplitude $q \left( t \right)$.
On the other hand, the entire physical mass $m_s$ of the microsphere moves rigidly by a displacement $q$. The total resonating mass $m$ referred to the free end (which appears for instance in Eq. (7) of the main text) is then $m=m_e+m_s$. 

The fact that the cantilever is not a rigid-body poses a difficulty in computing the collapse strength $\eta$ defined in Eq.~(3) of the main text, which can be determined only for rigid-body motions (strictly speaking, the center-of-mass master equation (2) is well-defined only for a rigid-body system). There is no easy way to cope with this situation, other than doing a numerical analysis of the full problem, or approximating the real motion with an appropriate rigid body motion. We choose the second approach, since the final bound on $\lambda$ is significant at level of order of magnitude. 

Therefore, we mimick the real motion of the cantilever with a cuboid which moves up and down rigidly, together with the sphere, with amplitude $q \left( t \right)$. According to the argument provided above, the effective cantilever motional mass which is rigidly moving by $q \left( t \right)$ is given by the fraction $\beta<1$ of the total cantilever mass. We then assume that the rigid cuboid mimicking the cantilever's motion has the length $R_1=\beta L$ along the $x$-axis, while the size along the other two directions (not affected by the elastic motion) is the same as that of the real cantilever. Notice that this choice is equivalent to approximate the real modal shape with a unit-step function modal shape, as shown in Fig.~\ref{fig:modeshape}, with same effective mass.

%%%%%%%%%%%%%%%%%%%%%%%%%%%%%%%%%%%%%%%%%%%%%%%%%%%%%%%%%%%%%%%%%%%%%%%%%%%%%%%%%%%%%%%%%%%%%%%%%%%%%%%%%%%%%%%
\section{CSL collapse strength}
\begin{figure}[ht!]
\includegraphics[width=15cm]{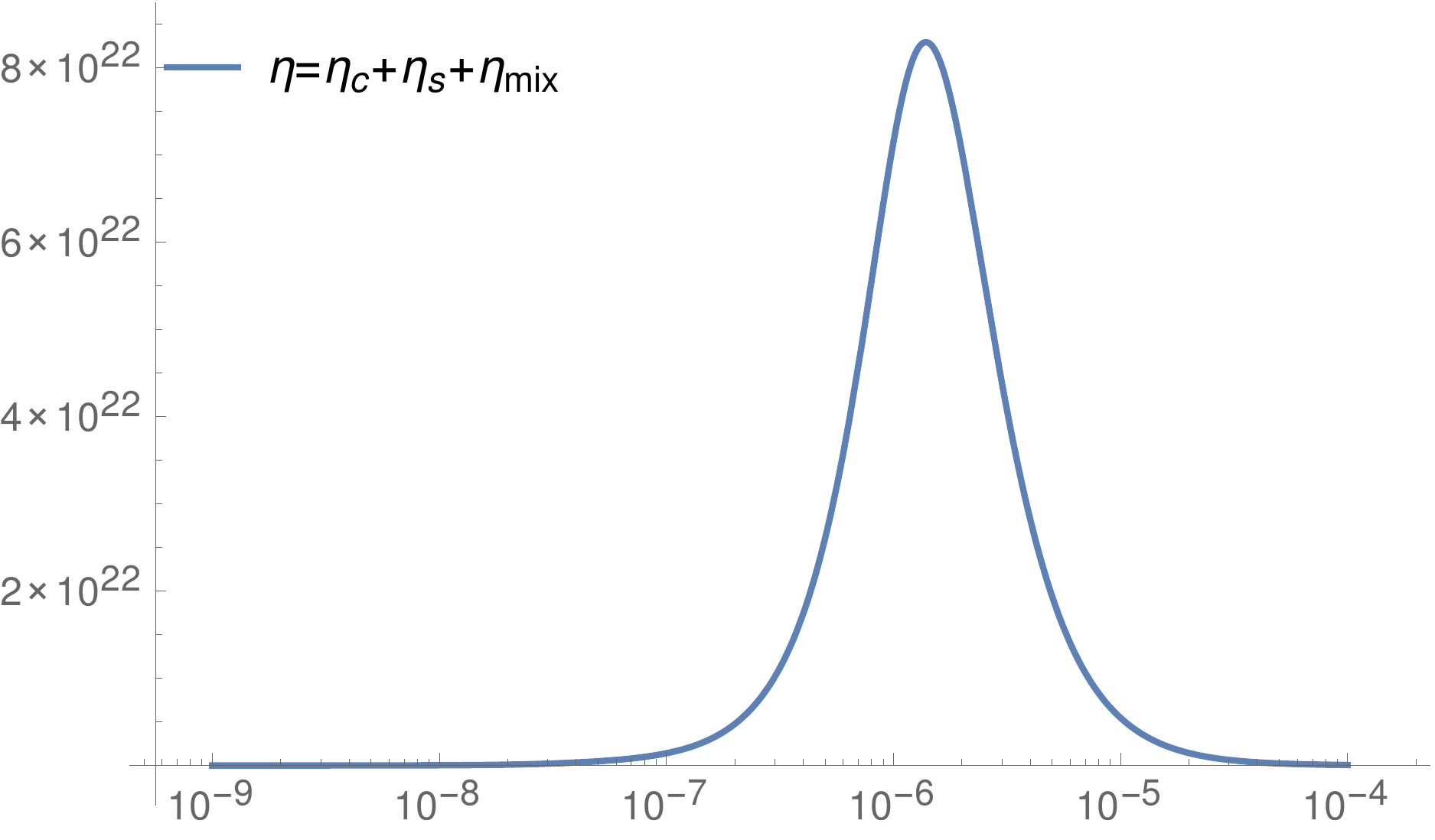}
\caption{The collapse strength $\eta$ in Eq.~\eqref{eta_T} as a function of $r_C$ where $\lambda=2.2 \times 10^{-17}$ s$^{-1}$ and all system's parameters are those given in the main text of the paper.}  \label{fig:eta1}
\end{figure}

%%%%%%%%%%%%%%%%%%%%%%%%%%%%%%%%%%%%%%%%%%%%%%%%%%%%%%%%%%%%%%%%%%%%%%%%%%%%%%%%%%%%%%%%%%%%%%%%%%%%%%%%%%%%%%%
The CSL collapse strength is given by Eq.~(3) of the main text:
\begin{align}
\label{eta1}
\eta &=\frac{(4\pi)^{\frac32}\,\lambda\,r_C^3}{m_0^2}\,
\int\frac{\D^3{\bf k}}{(2\pi)^3}\,k^2_z\,e^{-{\bf k}^2r_C^2}\,|\tilde{\varrho}({\bf k})|^2
\end{align}
with ${\bf r}=(x,y,z)$, ${\bf k}=(k_x,k_y,k_z)$, and $\tilde{\varrho}({\bf k})=\int\D^3{\bf r}\,e^{i{\bf k}\cdot{\bf r}}\,\varrho({\bf r})$ is the Fourier transform of the mass density.
Notice that [$\eta$]=m$^{-2}\cdot$s$^{-1}$. 
Given Fig.~1(a) of the main text, and considering the discussion in the previous section, the hybrid system can be described as a rigid-body system with the following density:
\begin{align}
\varrho({\bf r})=&\varrho_{\text{\tiny c}}({\bf r})+\varrho_{\text{\tiny s}}({\bf r})
\end{align}
with $\varrho_{\text{\tiny c}}({\bf r})$ the mass density of the cuboid, and $\varrho_{\text{\tiny s}}({\bf r})$ that of the sphere:
\begin{align}
\varrho_{\text{\tiny c}}({\bf r})=&\varrho_{\text{\tiny c}}\,\theta(x)\,\theta(R_1-x)
\,\theta(R_2/2-y)\,\theta(R_2/2+y)\,\theta(R_3/2-z)\,\theta(R_3/2+z)\\
\varrho_{\text{\tiny s}}({\bf r})&=\varrho_{\text{\tiny s}}\,
\theta(R-\sqrt{(x-(R_1-R))^2+y^2+(z-(R+R_3/2))^2}),
\end{align}
where $R_1=\beta\,L = 2.36\times10^{-6}$m, $R_2=w = 5\times10^{-6}$m, $R_3=d = 10^{-7}$m, $R = 2.25\times10^{-6}$m (see Fig.1(a) of the main text), $\varrho_{\text{\tiny c}}=2330\,\text{kg}\cdot\text{m}^{-3}$ is the uniform density of the cuboid, $\varrho_{\text{\tiny s}}=7430\,\text{kg}\cdot\text{m}^{-3}$ the uniform density of the sphere and $\theta(\cdot)$ is the Heaviside step function. 
Inserting the above densities into Eq.~\eqref{eta1}, one arrives at:
\begin{align}\label{eta_T}
\eta=\eta_{\text{\tiny c}}+\eta_{\text{\tiny s}}+\eta{\text{\tiny mix}}
\end{align}
with $\eta_{\text{\tiny s}}$ the collapse strength of the sphere:
\begin{align}\label{eta-s}
\eta_{\text{\tiny s}} &=
\frac{(4\pi )^2\,\lambda\,r_C^2\,R^2\,\varrho^2_{\text{\tiny s}}}{3m_0^2}\,
\left(
1-\frac{2r_C^2}{R^2}+e^{-\frac{R^2}{r_C^2}}\left(1+\frac{2r_C^2}{R^2}\right)
\right),
\end{align}
$\eta_{\text{\tiny c}}$ the collapse strength of the cuboid:
\begin{align}\label{eta-c}
\eta_{\text{\tiny c}} &=
\frac{32\lambda \,r^4_C\,\varrho_{\text{\tiny c}}^2}{m_0^2}
\left(1-e^{-\frac{R_3^2}{4r_C^2}}\right)
\left(e^{\frac{-R_2^2}{4r_C^2}}+
\frac{\sqrt{\pi} R_2}{2r_C}\text{Erf}\left(\frac{R_2}{2r_C}\right)-1\right)
\left(e^{-\frac{R_1^2}{4r_C^2}}+
\frac{\sqrt{\pi} R_1}{2r_C}\text{Erf}\left(\frac{R_1}{2r_C}\right)-1\right),
\end{align}
and $\eta{\text{\tiny mix}}$ the mixing of the two:
\begin{align}\label{eta-mix}
\eta_{\text{\tiny mix}} &=
2\frac{(4\pi)^{\frac32}\,\lambda\,r_C^3}{m_0^2}\,
\int\frac{\D^3{\bf k}}{(2\pi)^3}\,k^2_z\,e^{-{\bf k}^2r_C^2}\,
\text{Re}(\tilde{\varrho}_{\text{\tiny c}}({\bf k})\tilde{\varrho}^*_{\text{\tiny s}}({\bf k}))
\\&=
\frac{2\lambda}{m_0^2}\,\iint\,\D^3{\bf r}\,\D^3\mathbf{r}'\,
\exp\left(-\frac{|\mathbf{r}-\mathbf{r}'|^2}{4r_C^2}\right)\,
\frac{\partial\varrho_c({\bf r})}{\partial{z}}\,\frac{\partial\varrho_s({\bf r}')}{\partial{z'}}
\\&=
\frac{2\lambda\varrho_s\varrho_c\,R^2}{m_0^2}\,\int_0^{R_1}\D x\,\int_{-\frac{R_2}{2}}^{\frac{R_2}{2}}\D y\,
\int_{-\frac{R_3}{2}}^{\frac{R_3}{2}}\D z\,\int_0^\pi\D\theta\,\int_0^{2\pi}\D\phi\,
\sin\theta\cos\theta\left(\frac{R(1+\cos\theta)+R_3/2-z}{2r_C^2}\right)\times
\\&~~~~~~~~~~\nonumber
\exp\left[-\frac{(R\sin\theta\cos\phi+R_1-R-x)^2+(R\sin\theta\sin\phi-y)^2+(R\cos\theta+R+R_3/3-z)^2}{4r_C^2}\right].
\end{align}
Fig.~\ref{fig:eta1} shows the value of $\eta$ as a function or $r_C$. In Figs.~S3-S6 we single out the contribution  $\eta_{\text{\tiny s}}$ of the sphere, $\eta_{\text{\tiny c}}$ of the cuboid and the mixing term $\eta_{\text{\tiny mix}}$, respectively. As we can see, for $r_C \geq R, R_j$, the contribution of the sphere is dominant. 

To understand this behaviour, let us consider two limiting cases for  $\eta_{\text{\tiny s,c}}$ as given in Eqs.~\eqref{eta-s} and~\eqref{eta-c}.
For $r_C < R_j,R \sim 10^{-7}$m, one finds:
\begin{equation}
\begin{aligned}
\eta_{\text{\tiny c}}\approx\frac{8\pi \lambda R_1R_2\varrho^2_{\text{\tiny c}}}{m_0^2}\,r_C^2
=1.3\times10^{35}\,r_C^2;
~~~~~~
\eta_{\text{\tiny s}}\approx\frac{(4\pi)^2\lambda R^2\varrho^2_{\text{\tiny s}}}{3m_0^2}\,r_C^2
=1.2\times10^{35}\,r_C^2.
\end{aligned}
\end{equation}
Due to the geometry and mass density of the objects in our specific situation, the two contributions are of the same order of magnitude. Note the quadratic dependence on $r_C$: by increasing the correlation length, the number of nucleons, which contribute coherently to the collapse, increases, making the effect stronger. For very small values of $r_C$ (comparable or smaller than interatomic distances), the  approximation of a rigid body with a uniform density breaks down, and our formulas cannot be used any more.

On the other hand, for $r_C > R_j,R \sim 10^{-6}$m, one gets:
\begin{equation}
\begin{aligned}
\eta_{\text{\tiny c}}\approx\frac{\lambda}{2r_C^2}\left(\frac{R_1R_2R_3\varrho_{\text{\tiny c}}}{m_0}\right)^2
=3\times10^{9}\,r_C^{-2};
~~~~~~
\eta_{\text{\tiny s}}\approx\frac{\lambda}{2r_C^2}\left(\frac{\frac{4\pi}3R^3\varrho_{\text{\tiny s}}}{m_0}\right)^2
=5\times10^{11}\,r_C^{-2}.
\end{aligned}
\end{equation}
Now the contribution of the sphere to the collapse strength is largely dominant. This is because in this limit only the total number of nucleons becomes important; and in our case the sphere has more nucleons than the cuboid. Note  the quadratic dependence on the number of nucleons: all of them contribute coherently to the collapse. Note also the inverse quadratic dependence on $r_C$: for larger and larger values of the correlation length, there is no further coherent contribution to the collapse,  which then becomes weaker.

\begin{figure}[ht!]
\includegraphics[width=15cm]{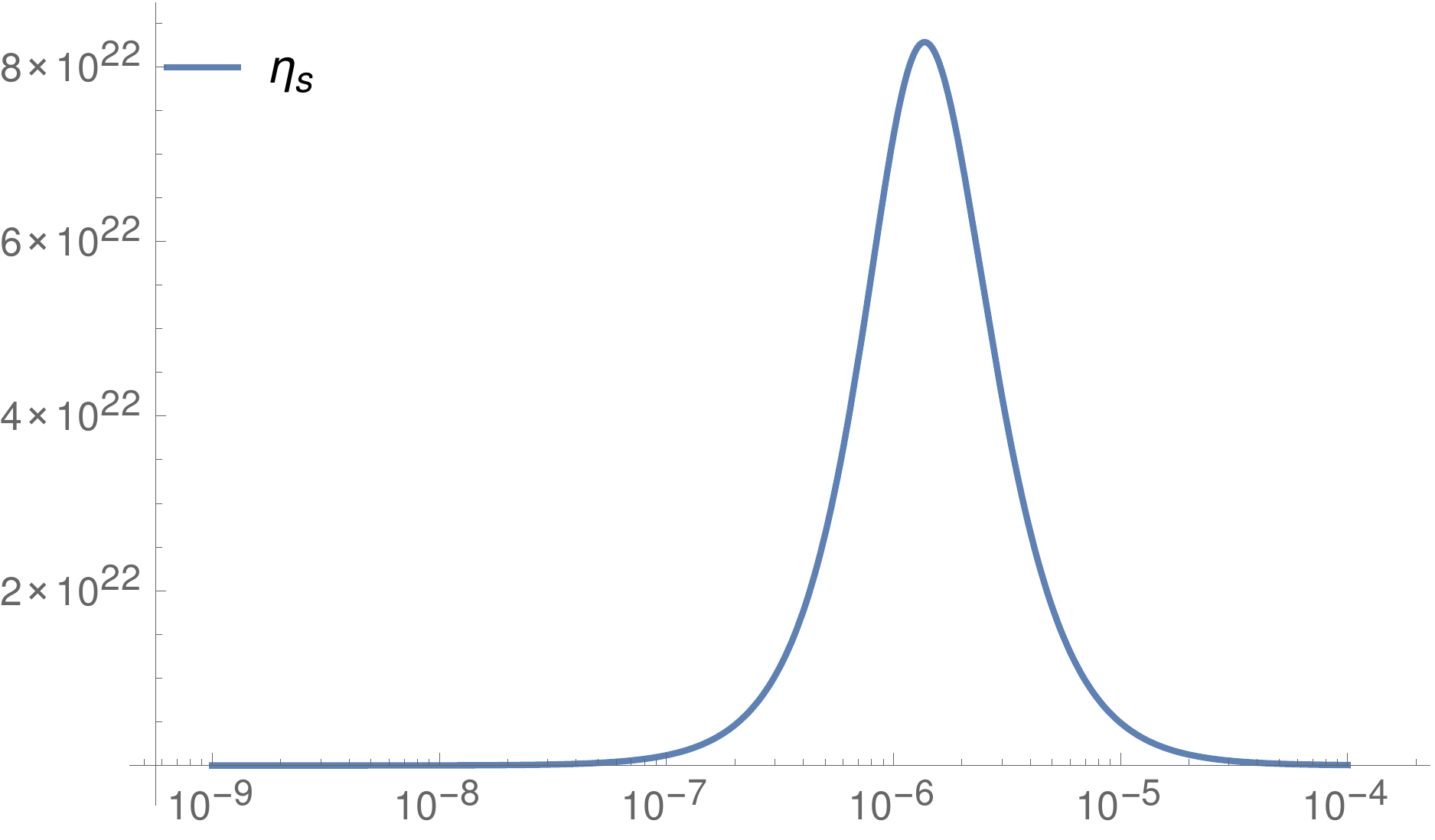}
\caption{The collapse strength $\eta_s$ in Eq.~\eqref{eta-s} as a function of $r_C$ where $\lambda=2.2 \times 10^{-17}$ s$^{-1}$ and all system's parameters are those given in the main text of the paper.}  \label{fig:eta2}
\end{figure}

\begin{figure}[ht!]
\includegraphics[width=15cm]{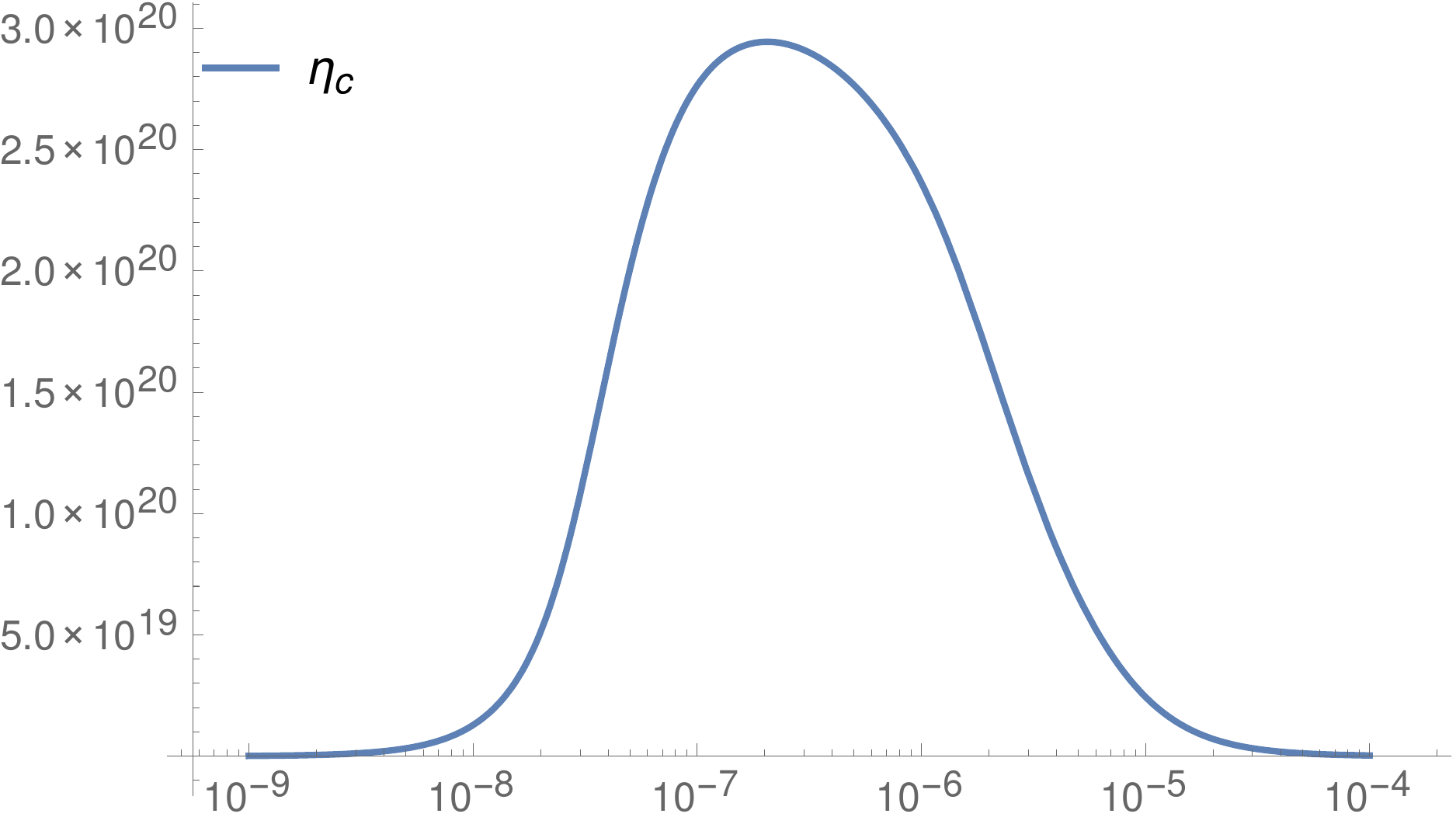}
\caption{The collapse strength $\eta_c$ in Eq.~\eqref{eta-c} as a function of $r_C$ where $\lambda=2.2 \times 10^{-17}$ s$^{-1}$ and all system's parameters are those given in the main text of the paper.}  \label{fig:eta3}
\end{figure}

\begin{figure}[ht!]
\includegraphics[width=15cm]{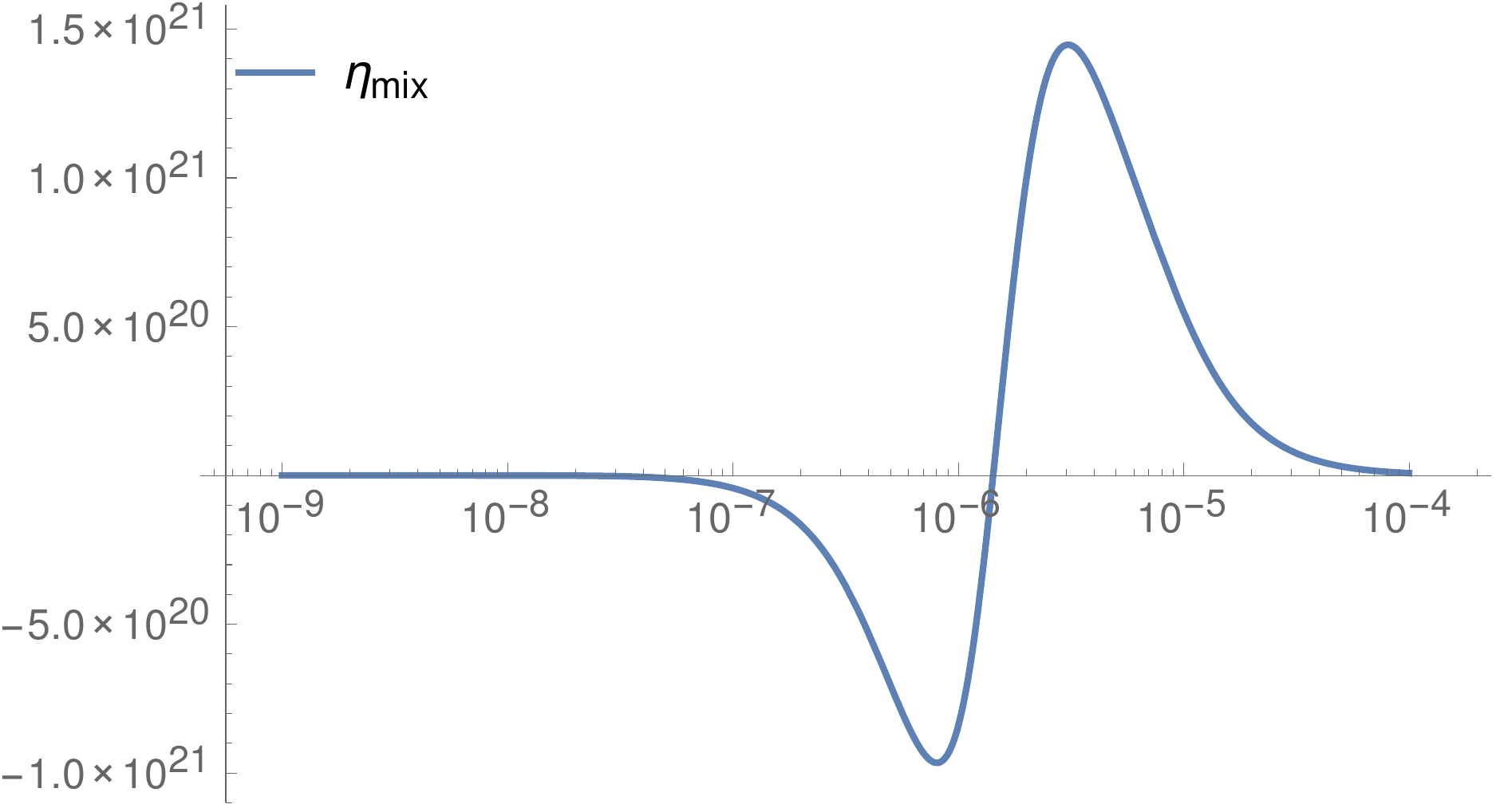}
\caption{The collapse strength $\eta_\text{mix}$ in Eq.~\eqref{eta-mix} as a function of $r_C$ where $\lambda=2.2 \times 10^{-17}$ s$^{-1}$ and all system's parameters are those given in the main text of the paper.}  \label{fig:eta4}
\end{figure}

\begin{figure}[ht!]
\includegraphics[width=15cm]{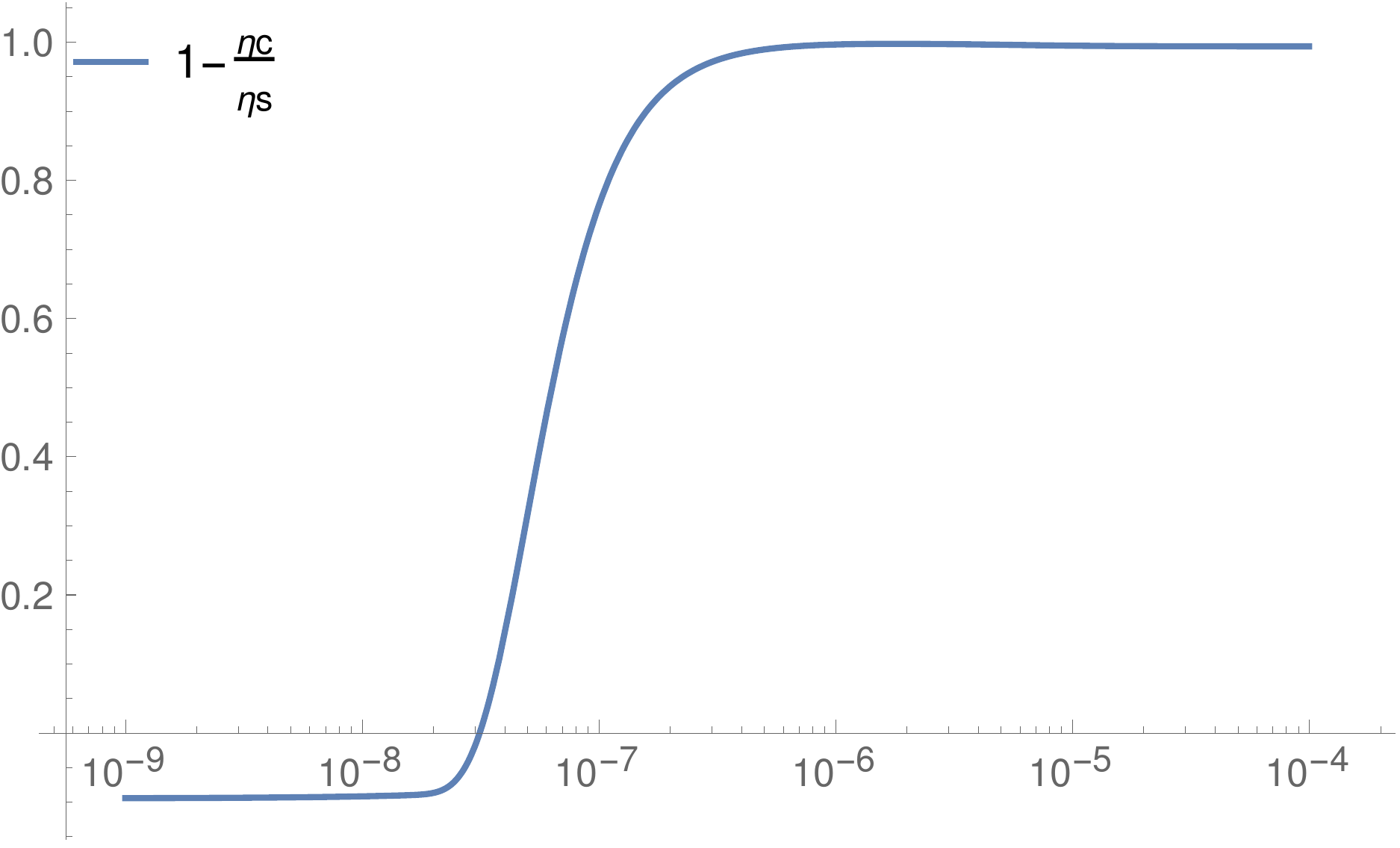}
\caption{The difference in collapse strengths for sphere and cuboid in Eqs.(\ref{eta-c},\ref{eta-s}) as a function of $r_C$ where $\lambda=2.2 \times 10^{-17}$ s$^{-1}$ and all system's parameters are those given in the main text of the paper.}  \label{fig:eta5}
\end{figure}

\end{document}